\documentclass[pageno]{jpaper}

\usepackage[normalem]{ulem}
\usepackage{amsmath}

\begin{document}

\title{
Effectiveness of Variable Distance Quantum Error Correcting Codes}
\author{Salonik Resch\\resc0059@umn.edu 
\and 
Ulya R. Karpuzcu\\
ukarpuzc@umn.edu}

\date{\hspace{6cm}University of Minnesota, Twin Cities}
\maketitle

\thispagestyle{empty}
\begin{abstract}
    Quantum error correction is capable of digitizing quantum noise and increasing the robustness of qubits. Typically, error correction is designed with the target of eliminating all errors - making an error so unlikely it can be assumed that none occur. In this work, we use  statistical quantum fault injection on the quantum phase estimation algorithm to test the sensitivity to quantum noise events. Our work suggests that quantum programs can tolerate non-trivial errors and still produce usable output. We show that it may be possible to reduce error correction overhead by relaxing tolerable error rate requirements. In addition, we propose using variable strength (distance) error correction, where overhead can be reduced by only protecting more sensitive parts of the quantum program with high distance codes.

\end{abstract}

\section{Introduction}
Physical quantum noise is hard to model accurately as it is highly complex and can have counter-intuitive impacts \cite{erhard2019characterizing}. Additionally, the noise present in a physical system can change over time \cite{tannu2019not}, making it difficult to properly characterize it via benchmarking procedures \cite{wilson2020just}.

Quantum error correction (QEC) groups collections of physical qubits togher to represent one logical qubit. The logical qubit is much more resilient to noise than the individual physical qubits it is made of. An additional useful property of QEC is that it restricts the quantum noise \footnote{Assuming the physical noise is sufficiently low to enable QEC to function properly}, forcing it to take on a restricted set of noise events, eg. X and Z errors. This allows us to accurately model the quantum noise acting on logical qubits with just X and Z errors \cite{steane1998introduction}.

For example, physical quantum noise can manifest as slight over- or under-rotations of individual qubits \cite{bravyi2018correcting}. The angle of over- or under- rotation can be any amount. In some sense, the noise is analog in nature. QEC involves the measurement of ancilliary qubits, which introduces non-unitary transformations of the quantum state (while leaving the computational quantum state intact). The measurement forces a decision on the error. Either the qubit snaps back to desired state, removing the noise, or it causes a complete noise event, flipping the state of the qubit. This is effectively a digitization of the noise into a set of errors. This digital nature of the errors allows them to be detected and corrected. 

QEC can only detect and correct errors on physical qubits if the physical noise rates are sufficiently low. If a sufficient number of noise events occur on multiple physical qubits simultaneously, this may result in an error which is either undetectable or uncorrectable - a logical error. QEC is typically designed with the target of removing logical noise events entirely. If a QEC code can tolerate a large number of simultaneous physical noise events, it becomes very unlikely that sufficient number of simultaneous noise events will occur. The minimum number of simultaneous error a QEC code tolerate is called the \emph{code distance}. 

Unfortunately, the cost of creating large distance QEC codes is very high. The number of physical qubits required for each logical qubit grows quadratically with the code distance, potentially reaching thousands of physical qubits for each logical qubits \cite{fowler2012surface}, even for reasonable noise rates. The resources quickly grow beyond what can realistically be provided by the available hardware.

In this work we suggest reducing the code distance and risk allowing some logical errors to occur. This can significantly reduce the resources required to build a (nearly) fault-tolerant computer. If the quantum application can tolerate some amount of error (still produces the correct output with high probability), such a strategy will prove effective. There are two methods to attempt
\begin{enumerate}
    \item Lowering the code distance overall (on all logical qubits at all times)
    \item Selectively lowering the code distance for a subset of the qubits at specific times
\end{enumerate}

In both cases, performing statistical fault injection with quantum applications will allow us to accurately estimate the probability of success, despite the existence of errors. Method 2 is enabled because Surface Codes \cite{fowler2012surface} allow us to change the number of physical qubits which are dedicated to each logical qubit. Hence, statistical fault injection can allow us to detect more and less sensitive regions of the application, and dedicate the resources more efficiently based on need. 

\vspace{-.2cm}
\section{Background}

Improving the reliability of quantum programs considering 
physical qubit characteristics
is a well studied problem. Numerous papers
such as \cite{tannu2019not} explored application mapping strategies considering variation in the reliability of physical qubits, while others focused on minimizing the number of gate evaluations to reduce the exposure time to noise \cite{dou2020new,li2019tackling}. 
Statistical fault injection based studies of program sensitivity
to physical noise, to optimize gate scheduling and qubit placement, also exist \cite{resch2020day}. An important distinction of our study from this lineage of work is that we 
operate at the logical qubit level rather than the physical.
%
Noise can be modeled more accurately at the logical level as QECC effectively forces noise to manifest as X and Z errors. This is in stark contrast to modeling noise at the physical level, which involves many different models. Worse, each noise model can result in a different outcome \cite{resch2019benchmarking}, and the noise in a physical experiment changes over time \cite{wilson2020just}.
{As a result, statistical fault injection at the physical level can produce contradicting conclusions depending on the noise model used \cite{resch2020day}.} 
%
At the same time, we primarily focus on fine-tuning QECC distance according to algorithmic needs, as opposed to common noise mitigation strategies at the physical level which focus on application mapping/scheduling.

\vspace{-.2cm}
\subsection{{Surface Codes}}
Surface codes are a promising form of QECC. They logically arrange qubits into a two-dimensional lattice, and only require interactions between nearest neighbors \cite{fowler2012surface,litinski2019game}, allowing the qubits to remain in place. 
This makes them easy to use with modern quantum computers, such as superconducting quantum computers, which have stationary qubits and only allow interactions between physically adjacent qubits. 
Surface codes can be conceptually visualized as ``patches'' of physical qubits, where patches can 
move and 
interact
with each other by performing operations and measurements on the qubits \cite{litinski2019game}. This is referred to as \emph{lattice surgery}, and represents the state of the art \cite{litinski2019game,horsman2012surface}. 
A surface code 
of code distance $d$ requires $d^2$ physical qubits per logical qubit. Logical gates on logical qubits require roughly $d$ time cycles \cite{litinski2019game}. Changing $d$ for each qubit also roughly takes $d$ cycles.

\vspace{-.2cm}
\subsection{{Gate Decomposition}}
A qubit can be logically represented as a point on the unit sphere (called Bloch sphere), where operations (gates) on it become rotations around different axis. $R_x(\theta)$, $R_y(\theta)$, and $R_z(\theta)$ gates correspond to rotations around the $x$, $y$, and $z$ axes by an arbitrary angle $\theta$. Such rotation gates are often used when operating quantum computers without QEC, where logical gates correspond directly to physical rotations on individual qubits. The specific gate set available 
depends on the specific machine, but generally precise rotations around at least two axes are typically available. However, such rotations do not work directly with surface codes (or most QECC). When the physical qubits are grouped together to form a 
logical qubit, only a specific set of gates are possible \cite{fowler2012surface}. In this work we use the Clifford+T gate set, the most widely used universal gate set which can also be used on surface codes. {It includes, among others, X, Y, Z, H, S, and T gates, which all correspond to rotations by $\pi$, $\pi/2$, or $\pi/4$ around various axes. Two-qubit variations of these gates exist, where a gate is performed on a \emph{target} qubit only if the \emph{control} qubit is in a specfic (i.e., the $\ket 1$) state. For example, the CNOT gate is a controlled X gate. A controlled phase gate is a controlled $R_z$ gate.}

Logical versions of $R_x(\theta)$, $R_y(\theta)$, and $R_z(\theta)$ gates are required for many algorithms, hence it is still necessary to implement them. To perform them on surface codes, they are \emph{approximated} with sequences of gates. For this, we use the 
 \emph{gridsynth} algorithm \cite{selinger2012efficient,gridsynth}, which converts $R_z(\theta)$ into sequences of $X$, $H$, $S$, and $T$ gates. The length of the sequence depends on the angle of rotation, $\theta$, and the precision to which we need to approximate it. For example, a rotation by $\pi/3$ can be approximated by 
 \begin{equation}
     R_z(\pi/3) \approx HSTHSHTHSHTST
 \end{equation}
 
 {There is also a trade-off between the sequence length and the achieved accuracy. Shorter sequences take less time to perform but 
 can lead to errors in the program
 (even in the absence of noise) due to the higher degree of approximation. We tested different sequence lengths when performing a noiseless simulation of a benchmark quantum algorithm (which we discuss in Section \ref{sec:sensitivity}). As an example, based on these experiments, we observe that
 for an average sequence length 
 of 
 34 gates per rotation, the correct output was produced only 50\% of the time. Increasing the sequence length to an average of 44 gates increased the probability of correct output to 98.5\%.}
 
 {Luckily, it is sufficient to approximate $R_z$ rotations, as any quantum operation $U$ can be decomposed into a set of three $R_z$ rotations (using three angles $\beta$, $\gamma$, and $\delta$) and H gates:}
 \begin{equation}
U = R_z(\beta)\;R_x(\gamma)\;R_z(\delta) =R_z(\beta)\;H\;R_z(\gamma)\;H\;R_z(\delta)
\end{equation}

In our case study, without loss of generality, we use standard Clifford+T gates, however, we note that further optimizations exist to tailor the operations specifically for surface codes \cite{litinski2019game}.

\subsection{Latency vs. Reliability Trade-Off}
\label{sec:tradeoff}
Reducing the QECC 
distance (i.e., making QECC weaker), by construction 
reduces the overhead, and hence, the time it takes to complete the program. 
On the other hand, a smaller code distance reduces the probability of success. 
%
Most quantum algorithms, even in the absence of noise, produce the correct result with some probability. 
Therefore, the algorithm 
has to be run multiple times before the correct answer is produced statistically. 
The number of repetitions depends on the probability of a successful trial (PST). 
${1}/{PST}$ runs 
are required to get the correct result, on average. Hence, \emph{mean time to success} is a key metric of interest. If the algorithm has a latency of $L$, the mean time to success 
becomes 
$L/PST$.
%
Let $L_{weak}$ and $PST_{weak}$ denote the latency and probability of success of the target algorithm using a weaker (i.e., smaller distance) QECC; where $L$ and $PST$ denote the latency and probability of success of the target algorithm using the default strength (i.e.,  distance) QECC.
\begin{equation}
    \begin{aligned}
        L_{weak} &< L\\
    PST_{weak} &< PST
    \end{aligned}
\end{equation}
applies, and 
a weaker QECC would only work if 
\begin{equation}
    \frac{L_{weak} }{ PST_{weak} }< \frac{L}{ PST}
\end{equation}
In the following, we will quantitatively analyze this trade-off.
\subsection{Sensitivity to Noise in Time and Space}
\label{sec:sensitivity}
\begin{figure*}
    \centering
    \includegraphics[width=0.7\textwidth]{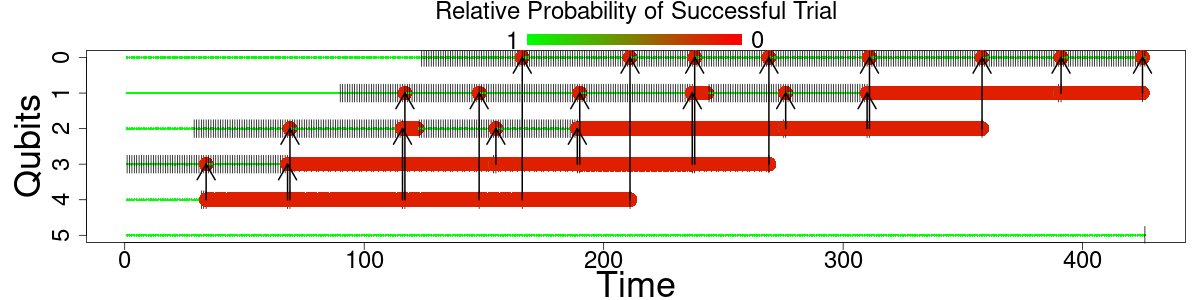}
    \caption{Noise sensitivity heatmap of a 6-qubit QPE alrogithm. Thin green lines represent insensitive;
    thick red lines, sensitive regions. Vertical lines correspond to single qubit gates. Arrows between qubits indicate two qubit interactions (i.e., gates). Horizontal lines terminate when the last operation on the respective qubit is complete, i.e., when the qubit is ready to be measured.   
    Qubit 5 has no operations on it after state preparation. {In our noise model, an error on a two-qubit gate affects both qubits.} }
    \label{fig:heatmap}
    \vspace{-.1cm}
\end{figure*}

The impact of a logical error on the success of a quantum program depends on 
when 
(at which cycle in the execution) and 
where
(in which qubit) 
the error occurs \cite{resch2020day}. 
We use 
\emph{statistical fault injection} to characterize this behavior, where we 
inject
errors at different locations at different times and 
track the
propagation to the output of the program. Since we are working with QEC, we can assume that errors are restricted to X and Z errors (Y errors are a combination of X and Z errors). As a representative case study, we profile 
the Quantum Phase Estimation (QPE) algorithm, which incorporates  
the inverse Quantum Fourier Transform (QFT) as the main computational kernel. QPE is representative of algorithms a fault-tolerant quantum computer (i.e., a quantum computer which features QEC) is likely to run, and 
has important applications in quantum chemistry \cite{lanyon2010towards}. {The input to QPE is a quantum state which has a phase angle difference between its constituent qubits. QPE detects this difference and produces a bitstring representing the angle. 6 qubits are large enough to show trends but small enough that visualization remains manageable. Patterns observed for 6 qubits hold for implementations of 
different qubit counts.}

To produce the input quantum state, we perform noiseless state preparation. A sequence of controlled-phase gates are performed with {qubits 0-4 as control and qubit 5 as the target. A noiseless QPE implementation acting on this state produces a single output bitstring with high probability} ($> 98\%$). {This eases validation of the output.} 

To guarantee statistical significance, we run many individual fault injection experiments.
In each experiment we inject 
an error 
to a single gate. We use Qiskit's \cite{cross2018ibm} depolarizing noise model with the error rate set to 100\% {(an error \emph{always} occurs)}, where X, Z, and Y (both X and Z) errors are all equally likely. We evaluate the quality of the output by comparing the probability of successful trial (PST) 
to that of the noiseless output. 
As we set the input state so that there is only one correct answer, this metric 
becomes:
\begin{equation}
    Relative\; PST = \frac{PST_{noisy}}{PST_{ideal}}
\end{equation}
Figure \ref{fig:heatmap} depicts the relative PST as a heatmap, to help visually
inspect
the significance of the error at each location and at each point in time. {The x-axis is time; the y-axis, space (i.e., different qubits).}
Thick, red lines indicate sensitive regions. Notably, qubits become more sensitive after they act as the control qubit of a two-qubit gate. {Intuitively, this is because the two-qubit gate entangles the qubits, merging two smaller states into a single, larger state.} Once entangled in this manner, qubits remain sensitive until the end of the program.

\vspace{-.1cm}
\section{Variable Strength QEC}
Knowledge about the underlying noise sensitivity of a quantum program, both in time and space, makes 
variable strength (i.e., distance) QEC possible, where different logical qubits 
get protected by 
different levels (i.e., different code distances $d$) of QEC  at different times. If a logical qubit is less susceptible to logical noise at a given point in time, it can survive with lighter weight QEC than more susceptible qubits.

We start by acknowledging a fundamental limitation to this idea. The logical error rate decreases \emph{exponentially} with $d$. The physical qubit count increases with $d^2$. The time overhead increases linearly with $d$. {\em Hence, we can save quadratic space and linear time, but risk exponential increases in error rates.} This suggests that variable-strength QEC can easily backfire if applied too aggressively.

\subsection{Success Rate as a Function of Code Distance}
We estimate the probability of logical error on a qubit, $P_L$ from the physical error rate, $p$, for code distance $d$ with the analytical formula provided by Fowler et al. for surface codes \cite{fowler2012surface}:
\begin{equation}
    P_L \approx 0.03 \times (p/0.0057)^{(d+1)/2}
\end{equation}

\begin{figure}
    \centering
    \includegraphics[width=.4\textwidth]{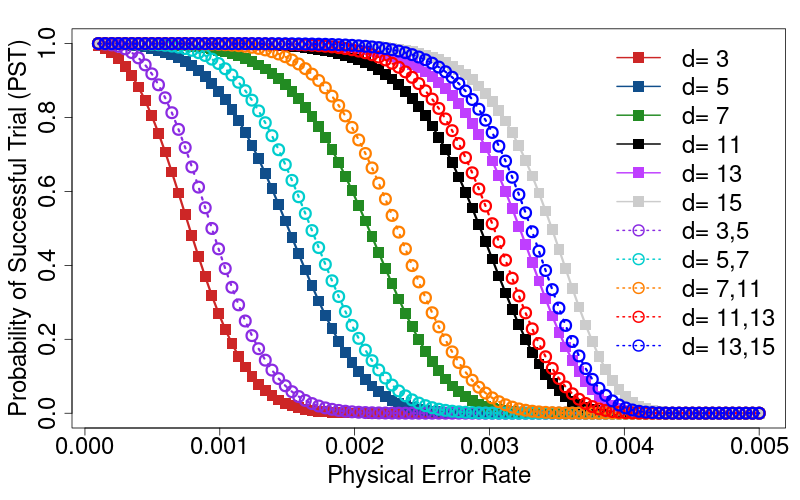}
    \vspace{-.2cm}
    \caption{PST of QPE for different (including variable) code distances over a range of physical noise rates.}
    \label{fig:pst}
    \vspace{-.2cm}
\end{figure}

The probability of successful trial (PST) is the probability of measuring 
the correct result at the end of the quantum program. With sufficiently high $d$, no logical errors would occur, and PST would be 1 (for quantum programs that have a single correct output). As $d$ decreases, 
PST decreases exponentially until it hits nearly 0. Where this decay occurs is a function of the physical error rate. Figure \ref{fig:pst} shows the PST of 
QPE 
for different $d$ over a wide range of physical error rates. Higher $d$ values tolerate higher physical error rates by construction. 

Using information from 
Figure \ref{fig:heatmap},
we know which logical qubits are more sensitive to noise, which we leverage to designate a variable distance QECC. We assign higher $d$ to the bold, red regions of Figure \ref{fig:heatmap}.
Without loss of generality, we experiment with two-distance QECC:
For example, $3,5$ is a QECC which uses $d=3$ on less susceptible qubits and $d=5$ on more susceptible qubits. 
The dashed lines in Figure \ref{fig:pst} capture the variable (two-) distance QECC,
which achieve resilience in-between their constituent code distances.  

\vspace{-.3cm}
\subsection{Time to Solution}
\label{sec:timetosolution}
We next combine the sensitivity information from Section \ref{sec:sensitivity} 
with the 
latency of each logic operation for a given code distance, to estimate the time to solution. As noted in Section \ref{sec:tradeoff}, the time to solution corresponds to ${L}/{PST}$. 

We estimate $PST$ by finding the probability of two events:
\begin{enumerate}
    \item No logical error occurs
    \item A single logical error occurs, but the output is the same as in the case of no error
\end{enumerate}
These two probabilities combined provide a lower bound on $PST$. It is also possible that two or more logical errors occur and the output remains the same. However, counting these possibilities quickly becomes intractable because a total of $(N_{gates})^{N_{errors}}$ must be considered, where $N_{gates}$ is the number of quantum gates performed in the algorithm and $N_{errors}$ is the number of possible errors. Hence, in our analysis, 
two or more errors represent a failure. 

\begin{figure}
    \centering
    \includegraphics[width=.4\textwidth]{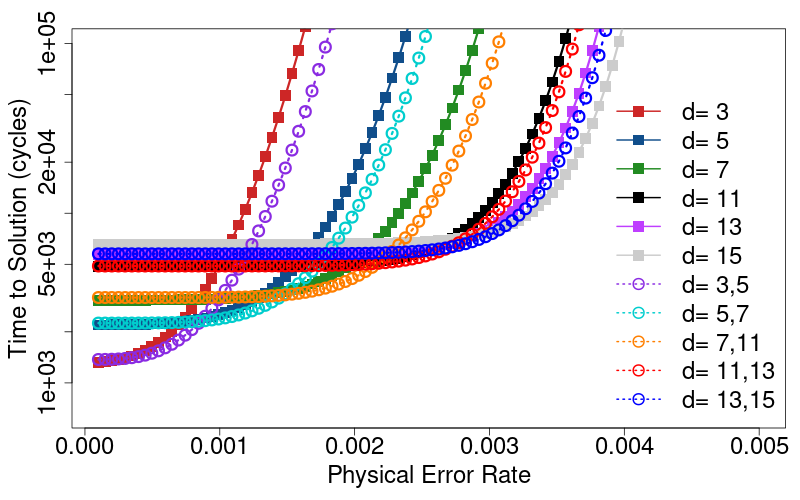}
    \vspace{-.2cm}
    \caption{The time to solution for different code distances. Variable distance codes tend to provide the fastest solution for a given error rate.}
    \label{fig:time}
    \vspace{-.2cm}
\end{figure}

A gate on a logical qubit with distance $d$ takes $d$ cycles to complete. Since 
the distance 
of each logical qubit is known throughout the program, we can easily find the corresponding latency, i.e.,  $L/PST$
which provides the time to solution as shown in Figure \ref{fig:time}.
It is noteworthy that the optimal code distance depends on the error rate. As intuition suggests, at low error rates lower code distances are preferable due to the lower overhead. However, as the error rate increases the codes begin to fail. Since error suppression is exponential, once the codes begin to fail, 
the reliability degrades dramatically and 
$P_{success}$ drops quickly. 
This necessitates a larger number of trials 
to obtain the correct solution, leading to a higher latency. 

Each variable distance code is optimal within a range of error rates. For example, the 
QECC $d=3,5$ is optimal at error rates where $d=3$ begins to fail. $d=3,5$ can provide a faster solution than $d=5$, until it breaks down and $d=5$ becomes necessary to tolerate errors. However, interestingly, variable distance codes tend to provide the best solution for most error rates. When $d=3,5$ fails, $d=5,7$ is already preferable to $d=5$, as shown in Figure \ref{fig:time}. 

\vspace{-.3cm}
\subsection{Practical Considerations}
While variable distance QECC looks promising, practical implementations can be challenging.
In expected superconducting quantum architectures, for which surface codes are most appropriate, logical qubits are arranged next to each other in a two-dimensional lattice \cite{javadi2017optimized}. {Non-uniform layouts incurred by expanding and contracting 
code distance (i.e., number of physical qubits per logical qubit)
can result in wasted physical qubits. Additionally, latency of such state-of-the-art quantum computers
is not limited by logic gates, but by the preparation of special quantum states required to perform specific operations (such as
magic state distillation for T gates} \cite{litinski2019game}).
Hence, improvements in the gate latency for logical qubits (as enabled by variable strength QECC) may not be significant.
Finally, quantum fault injection is tractable only for small circuits. Obtaining accurate sensitivity estimates for  larger circuits poses a challenge.
\vspace{-.3cm}
\section{Conclusion}
Our profiling analysis based on 
statistical fault injection shows that quantum programs can be relatively insensitive to isolated logical errors, and that variable distance QECC can reduce the time to solution by exploiting the spatio-temporal differences in noise sensitivity.
However, any decrease in code distance due to variable strength QECC comes with an exponential increase
in logical failure rates, which may eliminate the benefits if not carefully administered.

\bibliographystyle{plain}
\bibliography{references}

\end{document}